\documentclass[conference,letterpaper]{IEEEtran}

\usepackage[utf8]{inputenc} 
\usepackage[T1]{fontenc}
\usepackage[english]{babel}

\usepackage{dsfont}
\usepackage[cmex10]{amsmath}
\usepackage{amsfonts,amssymb,amssymb,mathrsfs,mathdots,mathtools}

\usepackage{xcolor}
\usepackage{graphicx}
\usepackage{float}
\usepackage{color}

\usepackage{stmaryrd}
\usepackage{verbatim}
\usepackage{paralist}   

\usepackage[colorinlistoftodos]{todonotes}

 \usepackage{url}

\pdfoutput=1
\newtheorem{theo}{Theorem}
\newtheorem{prop}{Proposition}
\newtheorem{lem}{Lemma}
\newtheorem{cor}{Corollary}
\newtheorem{rem}{Remark}
\newtheorem{defi}{Definition}

\newcommand{\vv}[1]{``#1''}

\makeatletter
\renewcommand*\env@matrix[1][*\c@MaxMatrixCols c]{%
  \hskip -\arraycolsep
  \let\@ifnextchar\new@ifnextchar
  \array{#1}}
\makeatother

\IEEEoverridecommandlockouts

\title{Remote Joint Strong Coordination \\and Reliable Communication}

\author{ 
  \IEEEauthorblockN{Giulia~Cervia, Tobias~J.~Oechtering, and Mikael~Skoglund}
  
  \IEEEauthorblockA{School of Electrical Engineering and Computer Science,
                    KTH Royal Institute of Technology,\\
                    Stockholm, Sweden,
                    \{cervia, oech, skoglund\}@kth.se}
  \thanks{This work was supported in part by the Swedish foundation for strategic
research and the Swedish research council.}
}

\begin{document}
\maketitle

 \begin{abstract} 
We consider a three-node network, in which  two agents wish to communicate over a noisy channel, while controlling the distribution observed by a third external agent.
We use strong coordination to constrain the distribution, and we provide a complete characterization of the \vv{remote  strong coordination and reliable communication} region.
 \end{abstract}


\section{Introduction}

Coordination was introduced in~\cite{cuff2009thesis} as a generalization of traditional information problems, and it is intended as a way to enforce a prescribed behavior and to align statistical information over a network. 
However, so far there are relatively few results on coordination with security constraints, with
the exception of~\cite{satpathy2014secure, satpathy2016secure} that looked at secure coordination with noiseless links.



In this paper we address for the first time the problem of \emph{remote joint strong coordination and reliable communication} with an outside observer and a noisy channel, which is depicted in Fig.~\ref{fig: shannonz}.
We study a three-node network model comprised of an information source and a noisy channel, in which two agents, an encoder and a decoder, have access to a common source of randomness. Moreover, we consider a third agent, Eve, who observes an output of the noisy channel (possibly different from the decoder's) but has no knowledge of the common randomness.

We propose a general template problem formulation which presents two different goals: the encoder needs to reliably convey a message to the decoder, while simultaneously forcing  Eve's observation.
We can think of two relevant scenarios that can benefit from coordination in this sense. As traditionally in  security and privacy, Eve can be a malicious eavesdropper, and, by remotely coordinating her observation, we control what the eavesdropper sees. Then, for example, we can impose perfect secrecy~\cite{bloch2011physical}, and have the message independent of  Eve's observation. The second possibility is that Eve is not an adversary, but a neighboring receiver, and we want to limit the interference created to Eve by the communication process, as studied for the weaker notion of empirical coordination in~\cite{blasco2014communication}.

We characterize the  remote strong coordination and reliable communication region. Moreover, we are able to characterize the region \begin{inparaenum}[i)]
\item when the distribution that the eavesdropper sees is independent of the message exchanged over the legitimate channel, 
\item when we want to control only the eavesdropper's marginal distribution.\end{inparaenum}

The rest of the document is organized as follows. 
Section~\ref{sec: prel} introduces the notation, while Section~\ref{sec: model} describes the model under investigation, and states the main results.
Finally, the achievability and converse proofs 
are found in Section~\ref{sec: proof} and Section~\ref{section converse} respectively.
\section{Notation}\label{sec: prel}

We define the integer interval $\llbracket a,b \rrbracket$ as the set of integers between $a$ and $b$.
Given a random vector $X^{n}\coloneqq$ $(X_1, \ldots, X_{n})$, we denote $X^{i}$ as the first $i$ components of $X^{n}$, and $X_{\sim i}$ as the vector $(X_j)_{j \neq i}$, $j\in \llbracket 1,n \rrbracket $, i.e., $X^{n}$ without the component $X_i$. 
Similarly, we denote  $X^n_{\llbracket 1, m\rrbracket}$ as the set of vectors $\{X_s^n |\,  s \in \llbracket 1,m \rrbracket \}$ and 
 $X_{\mathcal S}^n$,  as the set of vectors $\{X_s^n |\, s\in  \mathcal S,  \, \mathcal S\subseteq \llbracket 1,m \rrbracket \}$.
We indicate with $\mathbb V (\cdot , \cdot)$  the  \emph{total variation distance} (or \emph{variational distance}) between two distributions. 
We use $P_{A}^{\otimes n}$ for the \emph{i.i.d. product distribution} associated with $P_{A}$, and $Q_A$ for the \emph{uniform distribution} over $\mathcal A$. 
We denote with $\varepsilon$ a positive infinitesimal quantity which tends to zero as $n$ goes to infinity.

\section{System Model and Main Results}\label{sec: model}

\begin{center}
\begin{figure}[h!]
\centering
\includegraphics[scale=0.18]{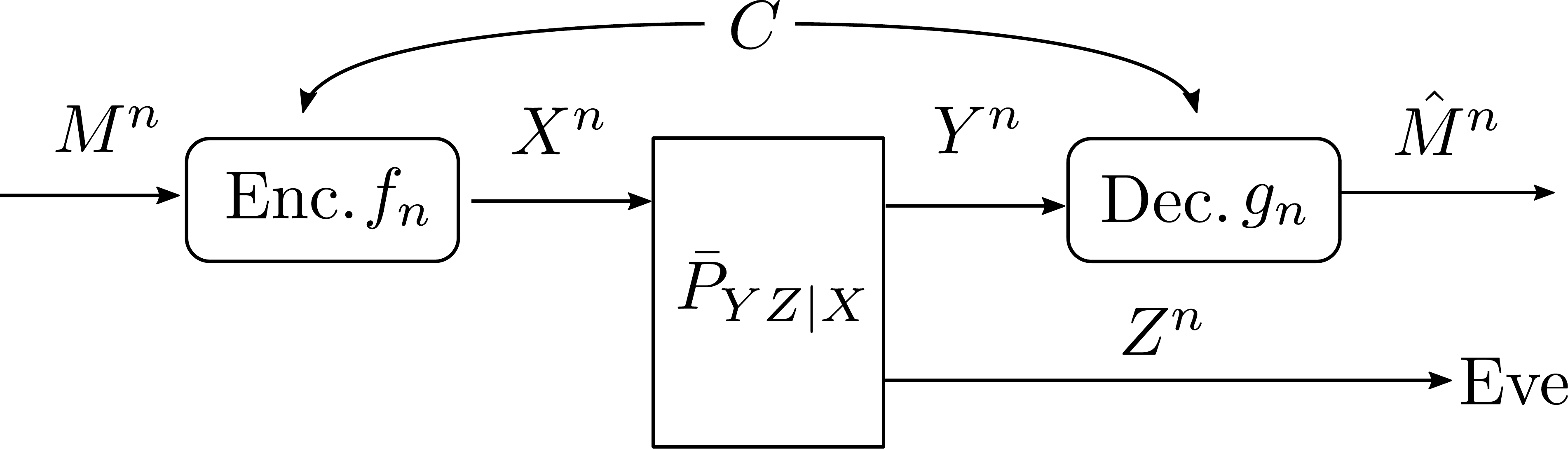}
\caption{Communication system: a transmitter wants to reliably send a sequence of uniform messages $(M_1, \ldots, M_{n})$ of total rate $R_{M}$ over a discrete memoryless channel $\bar P_{Y|X}$, while simultaneously controlling the distribution on $Z^n$ seen by Eve.}
\label{fig: shannonz}
\end{figure}
\end{center} 

\vspace{-10mm}
\subsection{System Model}\label{ssec:syst_model}
Suppose we are in the setting of Figure \ref{fig: shannonz}.  
The encoder wants to send a uniform message $M^n$ of rate $R_{M}= \log K^n/n= \log K$
over a DMC $\bar P_{Y|X}$, and encoder and decoder share a source of uniform randomness of rate $R_{C}$.
We want to think of the message $M^n$ 
as a sequence of \emph{uniform i.i.d. messages $(M_1, \ldots, M_{n})$}. where each $M_i$ is generated uniformly according to the distribution $Q_M$. To do that, we write the message $M^n$ as the integer $1+ \sum_{i=1}^n (M_i-1) K^{i-1}$ $\in \llbracket 1, K^n\rrbracket$, which is characterized  by the choice of the coefficients $M_i \in  \llbracket 1, K \rrbracket$, and we denote it with $M^n\coloneqq$ $(M_1, \ldots, M_{n})$.

The decoder exploits its knowledge of the  output of the channel and of common randomness $C$  to estimate correctly the message  $M^n$. 
At the same time,  we want  to remotely control the output of the DMC $\bar P_{Z|X}$ seen by the eavesdropper $Z^n$, by having the distribution $P_{Z^n|M^n}$ induced by the code  indistinguishable from a target i.i.d. distribution.
We formalize these requests in the following definition.
\vspace{1mm}
\begin{defi}\label{definitionscz}
Given a channel $\bar P_{YZ|X}$,
a triple $(\bar{P}_{Z|M},$  $R_{M},$ $R_{C})$ is achievable for \emph{remote strong coordination and reliable communication} if there exists a sequence $(f_n, g_n)$ of encoders-decoders with rate of common randomness $R_{C}$, such that, for every message $M^n$,
\vspace{-1mm}
{\allowdisplaybreaks\begin{subequations}
\begin{align}
&\lim_{n \to \infty} \mathbb V \left( P_{Z^{n}|M^n}, \bar{P}_{Z|M}^{\otimes n} \right) =0, \label{coordz}\\
& \lim_{n \to \infty} \mathbb P \,\{ \hat{M}^n \neq M^n \}=0, \label{m}
\end{align}\end{subequations}}where $P_{Z^n}$ is the joint distribution induced by the source, the channel, and the coordination code.
The \emph{remote strong coordination and reliable communication region}   $\mathcal{R}$ is the closure of the set of all achievable triples $(\bar{P}_{Z}, R_{M}, R_{C})$.
\end{defi}
\begin{rem}
Note that even though we define the message as a sequence of uniform messages, the problem is well-defined. In fact, if one considers a permutation of the message set, the set of achievable triplets does not change.\end{rem}

\subsection{Main Results}
In this section, we present the \emph{remote strong coordination and reliable communication} regions, first in the general case and then with secrecy constraints.
\subsubsection{General Case}

\vspace{1mm}

\vspace{1mm}
\begin{theo}\label{theogeneral}
Let  $\bar P_{YZ|X}$ be the given  channel parameter, then 
\begin{equation}
\mathcal R  \! \coloneqq \!
\begin{Bmatrix*}[l]
 (\bar P_{Z|M}, R_{M}, R_{C}) : \\
 \quad \bar P_{Z|M}=\sum_x P_{X} \bar P_{Z|MX}  \\
  \quad \exists \,W \sim \bar P_{W|M^nCXYZ} \mbox{ such that }\\
  \quad W- X-YZ\\
 \quad R_{M} \leq \max_{P_X}\! \min{\{ I(X;Y); H(XW|Z)\}} \\
 \quad R_{C} \geq I(W;Z)-I(W;Y)  \\
 \quad \lvert \mathcal W \rvert\leq  \lvert \mathcal X \times \mathcal Y \times \mathcal Z \rvert +1
\end{Bmatrix*}.\label{region gen}\end{equation}
\end{theo}

\vspace{1mm}
\subsubsection{Secrecy constraints}
Suppose  we not only wish  to control what the decoder sees, by strongly coordinating $P_{Z^n|M^n}$, but we also want to make sure that the message $M^n$, safely exchanged between the legitimate users, is independent of the observation $Z^n$: 
\vspace{-2mm}
\begin{equation}\label{secrecyz}
 \lim_{n \to \infty} I(M^n;Z^n)= 0.
\end{equation}Then, by adding the strong secrecy condition~\cite{elgamal2011nit} above to the conditions \eqref{coordz} and \eqref{m} of Definition~\ref{definitionscz}, the following result derives the coordination region.
\begin{prop}\label{secrecy region}
Let  $\bar P_{YZ|X}$ be the given  channel parameter, then the following region is the strong coordination region as defined in Definition~\ref{definitionscz}, with the strong secrecy condition~\eqref{secrecyz}:
\begin{equation}
  \mathcal R_{\text{ind}} \coloneqq \!\begin{Bmatrix*}[l]
 (\bar P_{Z|M}, R_{M}, R_{C}) : \\
 \quad \bar P_{Z|M}=\sum_x P_{X} \bar P_{Z|MX}  \\
  \quad \exists \, W \sim \bar P_{W|M^nCXYZ} \mbox{ such that }\\
  \quad W- X-YZ\\
 \quad R_{M} \leq \max_{P_X}\! \min{\{I(X;Y), H(X|Z)\}} \\
 \quad R_{C} \geq I(W;Z) - I(W;Y)   \\
 \quad \lvert \mathcal W \rvert\leq  \lvert \mathcal X \times \mathcal Y \times \mathcal Z \rvert +1
\end{Bmatrix*}\!.\label{regionz}\end{equation}
\end{prop}


\subsection{Control of the marginal distribution}
Suppose that, instead of considering $P_{Z^n|M^n}$, we make the weaker demand of controlling the distribution $P_{Z^n}$ observed by the eavesdropper. 
As we will see, this request has the advantage of having the rate $R_M $ upper bounded only by $I(X;Y)$, allowing the legitimate receiver to exploit the entire capacity of the channel, regardless of the other dependencies between random variables.
Then, the set of distributions and rates for which 
\begin{subequations}
\begin{align}
&\lim_{n \to \infty} \mathbb V \left( P_{Z^{n}}, \bar{P}_{Z}^{\otimes n} \right) =0, \label{coordz2}\\[1mm]
& \lim_{n \to \infty} \mathbb P \,\{ \hat{M}^n \neq M^n \}=0, \label{m2}
\end{align}\end{subequations}is characterized in the following result.

\begin{prop}\label{control z}
Let  $\bar P_{YZ|X}$ be the given  channel parameter, the capacity region for which~\eqref{coordz2} and~\eqref{m2} hold, is
\begin{equation}
\mathcal R_{Z^n}  \coloneqq 
\begin{Bmatrix*}[l]
 (\bar P_{Z}, R_{M}, R_{C}) : \\
 \quad \bar P_{Z}=\sum_x P_X \bar P_{Z|X}  \\
   \quad \exists W \sim \bar P_{W|M^nCXYZ} \mbox{ such that }\\
  \quad W- X-YZ\\
 \quad R_{M} \leq \max_{P_X} I(X;Y)\\
 \quad R_{C} \geq I(W;Z) - I(W;Y) \\
  \quad \lvert \mathcal W \rvert\leq  \lvert \mathcal X \times \mathcal Y \times \mathcal Z \rvert +1
\end{Bmatrix*}.\label{region control z}\end{equation}
\end{prop}
\vspace{1mm}
\begin{rem}
Note that $\mathcal R$ and $\mathcal R_{Z^n}$ do not live in the same space, since the space of probability distributions is different. Hence,  one is not contained in the other.
\end{rem}

\vspace{1mm}
\subsubsection*{Related work}
Note that both the constraint on $R_M$and  the achievable distributions are the same for the empirical coordination region for this setting, derived in~\cite{blasco2014communication}. 
However, strong coordination requires a positive rate of common randomness. This is coherent with the conjecture that, with enough common randomness,
the strong coordination region is the same as the empirical coordination region for any network setting~\cite{cuff2010}.

\section{Achievability}\label{sec: proof}

The achievability proofs have the following structure:
\begin{enumerate}
\item[\emph{1)}] An  i.i.d. random binning  scheme and a random coding scheme are presented;
 \vspace{1mm}
\item[\emph{2)}]\label{step2ach} Strong coordination of $(M^n, C, F, W^n, X^n, Y^n, Z^n)$, by showing that the two schemes have the same statistics;
 \vspace{1mm}
\item[\emph{3)}] Proof of the reliable communication problem;
 \vspace{1mm}
\item[\emph{4)}]\label{step4ach} Optimization of the schemes  to have coordination of the sequences of our interest only by reducing the rate of common randomness;
 \vspace{1mm}
\item[\emph{5)}] Summary of  the rate conditions.
\end{enumerate}

In particular, steps 2 and 4 use extensively the properties of the variational distance and the random binning techniques~\cite{yassaee2014achievability} here summarized. 
\vspace{1mm}

\begin{lem}[Properties of the total variation distance]\label{tv prop}
\begin{enumerate}[(i)]
 \item \label{cuff16} $\mathbb V (P_{A}, \hat P_{A}) \leq \mathbb V (P_{AB}, \hat P_{AB})$, see~\cite[Lemma 16]{cuff2009thesis};
 \item \label{cuff17} $\mathbb V (P_A, \hat P_A)= \mathbb V (P_AP_{B|A}, \hat P_A P_{B|A})$, see~\cite[Lemma 17]{cuff2009thesis};%
 \item \label{lem4y} if $ \mathbb V (P_{A} P_{B|A}, P'_{A} P'_{B|A})$ $ = \varepsilon$, then there exists $a \in \mathcal A$ such that $\mathbb V (P_{B|A=a},P'_{B |A= a} ) \leq 2 \varepsilon$, see~\cite[Lemma 4]{yassaee2014achievability}.
\end{enumerate}
\end{lem}

\paragraph*{Properties of random binning}
Let $(A_{\llbracket 1, N \rrbracket}, B)$ be discrete memoryless correlated sources distributed according to  $P_{A_{\llbracket 1, N \rrbracket}, B}$ on $\prod_{i=1}^N \mathcal A_i^n \times \mathcal B$. A distributed random binning $\varphi \coloneqq (\varphi_1, \ldots, \varphi_N)$ consists of a set of uniform random mappings $\varphi_{i}: \mathcal A_i^n \to \llbracket 1, 2^{nR_i}\rrbracket$, $i\in \llbracket 1,N\rrbracket$, and we denote $C_i  \coloneqq  \varphi_{i}(A_i^n)$.

The first result ensures that the binnings are almost uniform and almost independent from $B^n$.
\begin{theo}[$\mbox{\cite[Theorem 1]{yassaee2014achievability}}$]\label{th1y}
If for every $\mathcal S \subseteq \llbracket 1, N \rrbracket$, 
\begin{equation*}
\sum_{s \in \mathcal S} R_s < H(A_{\mathcal S}|B),
\end{equation*}then as $n$ goes to infinity,
\begin{equation*}
\mathbb{E}_{\varphi} [ \mathbb V(P_{C_{\llbracket 1, N \rrbracket} B^n} , Q_{C_{\llbracket 1, N \rrbracket} } P_{B^n} ) ] \to 0.
\end{equation*}
\end{theo}

\vspace{2mm}
Then, we use the following to ensure the independence of one binning only.
\begin{cor}[$\mbox{\cite[Corollary 1]{yassaee2014achievability}}$]\label{cor1y}
Let $\mathcal V$ an arbitrary subset of $\llbracket 1, N \rrbracket $. 
If for every $\mathcal S \subseteq \llbracket 2, N \rrbracket \setminus \mathcal V$, 
\begin{equation*}
R_1 + \sum_{s \in \mathcal S} R_s < H(A_1 A_{\mathcal S}|B A_{\mathcal V}),
\end{equation*}then as $n$ goes to infinity,
\begin{equation*}
\mathbb{E}_{\varphi} [ \mathbb V(P_{C_{\llbracket 1, N \rrbracket} B^n} , Q_{C_{1} } P_{ C_{\llbracket 2, N \rrbracket}  B^n} ) ] \to 0.
\end{equation*}
\end{cor}

\vspace{2mm}
Now, we state a lemma for  a special case of the Slepian--Wolf Theorem for recovering only one source $A_1^n$ from random bins $C_1, \ldots C_N$ and  side information $B^n$. 
\begin{lem}[$\mbox{\cite[Lemma 2]{yassaee2014achievability}}$]\label{lem2y}
If for every $\mathcal S \subseteq \llbracket 2, N \rrbracket$, 
\begin{equation*}
R_1 + \sum_{s \in \mathcal S} R_s > H(A_1 A_{\mathcal S}|B A_{\mathcal S^c}),
\end{equation*}then as $n$ goes to infinity,
\begin{equation*}
\mathbb{E}_{\varphi} [ \mathbb V(P_{A^n_{\llbracket 1, N \rrbracket}  B^n \hat A^n_1} ,  P_{ A^n_{\llbracket 1, N \rrbracket}  B^n} \, \mathds 1\{\hat A^n_1 = A^n_1\}) ] \to 0
\end{equation*}where $\hat A^n$ is the output of a  Slepian--Wolf decoder.
\end{lem}

\subsection{Achievability Proof of Theorem~\ref{theogeneral}}\label{sec: gen proof a}

\subsubsection{I.i.d. random binning and random coding scheme}\label{subsec: def random schemes}
We present two distributions, a random binning scheme to  synthesize the target i.i.d. distribution, and a random coding scheme to achieve the average performance.
\paragraph{Random binning scheme}\label{rb gen}
We consider  $W^{n}$ generated i.i.d. according to $\bar P_{W^n}$ $\coloneqq$ $\bar P_W^{\otimes n}$, and $X^n$ generated i.i.d. according to $\bar P_{X^n|W^n}$ $\coloneqq \bar P_{X|W}^{\otimes n}$.
Then, to all the sequences $\mathbf{w} \in \mathcal W^n$ we assign indices through the following uniform random binning: 
\begin{itemize}\setlength{\itemsep}{0.2em}
\item $\varphi_F: \mathcal{W}^{n} \to \llbracket 1,2^{nR_F} \rrbracket$, $F = \varphi_F(W^{n})$.
\end{itemize}Moreover, to all the sequences $\mathbf{x} \in \mathcal X^n$ we assign indices through the following uniform random binnings: 
\begin{itemize}\setlength{\itemsep}{0.2em}
\item  $\varphi_C: \mathcal{X}^{n} \to \llbracket 1,2^{nR_C} \rrbracket$,  where $C = \varphi_C(X^{n})$ represents the codebook;
\item $\varphi_{M}: \mathcal{X}^{n} \to \llbracket 1,2^{n R_{M}} \rrbracket$, where $M^n = \varphi_{M}(X^{n})$, represents the message.
\end{itemize}
The i.i.d. joint distribution induced by the binnings is:
{\allowdisplaybreaks
\begin{align}
\MoveEqLeft[1.5]
P^{\text{RB}}_{M^n C F W^n X^nY^n Z^n} \nonumber\\
\coloneqq& \bar P_{W^n} P_{F|W^n}^{\varphi_F}  \bar P_{X^n|W^n} P_{C|X^n}^{\varphi_C} P_{M^n|X^n}^{\varphi_M}  \bar P_{Y^nZ^n|X^n}\nonumber\\
 =&P^{\text{RB}}_{FM^n C} P^{\text{RB}}_{W^n|F}  P^{\text{RB}}_{X^n| W^n M^n C} \bar P_{Y^nZ^n|X^n}, \label{rb1}
\end{align}}where the pair of well-defined distributions $(P^{\text{RB}}_{W^n|F},$ $P^{\text{RB}}_{X^n| W^n M^n C})$ is used as an encoder in the random coding scheme.

\paragraph{Random coding scheme}\label{rb gen}
We consider the joint distribution induced by the code:
{\allowdisplaybreaks
\begin{align}
\MoveEqLeft[1.5]
P^{\text{RC}}_{M^n C F W^n X^nY^n Z^n} \nonumber\\
&\coloneqq \!\!Q_{M^{\!n}\!} Q_{C} Q_{F} P^{\text{RB}}_{W^n|F}  P^{\text{RB}}_{X^n| W^n\! M^n C} \bar P_{Y^n \! Z^n | X^n}\!,\label{rc1}
\end{align}}and we choose the encoders $P^{\text{RB}}_{W^n|F} $, and $P^{\text{RB}}_{X^n| W^n M^n C}$ as defined in~\eqref{rb1}.

\begin{center}
\begin{figure}[h!]
\centering
\includegraphics[scale=0.052]{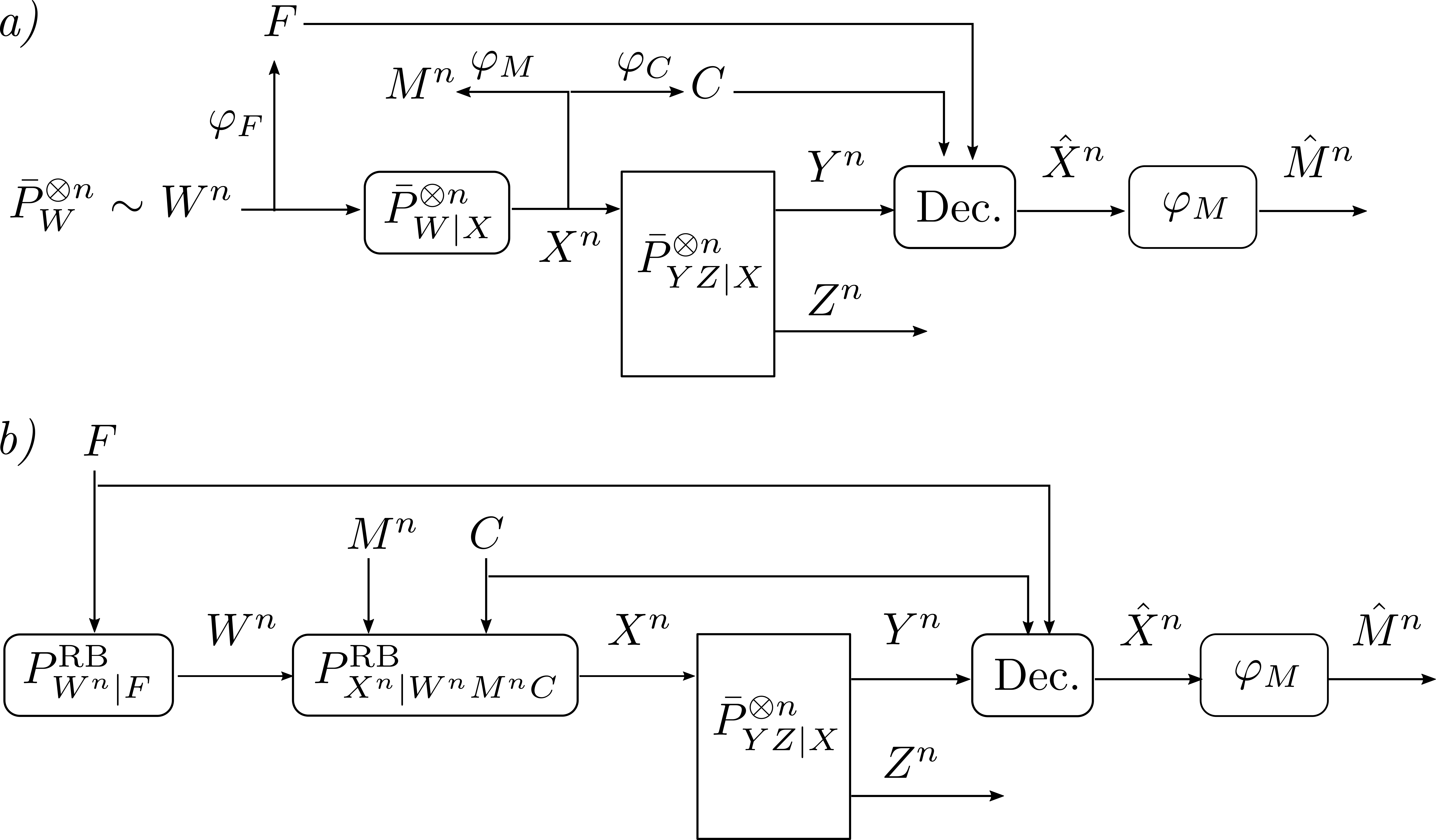}
\caption{Proposed schemes and channel code construction: \emph{a)} random binning scheme, \emph{b)} random coding scheme.}
\label{fig: schemes}
\end{figure}
\end{center} 

\subsubsection{Approximating the target distribution -- Strong coordination of $(M^n, C, F, W^n,X^n, Y^n, Z^n)$}\label{subsec: joint strong coord}
Observe that 
{\allowdisplaybreaks
\begin{align}
&\mathbb V (P^{\text{RB}}_{M^n C F W^nX^nY^n Z^n}, P^{\text{RC}}_{M^nCFW^n X^nY^n Z^n})\nonumber\\[1mm]
& \overset{\mathclap{\smash{(a)}}}{=}  \mathbb V(P^{\text{RB}}_{M^n C F W^nX^n}, P^{\text{RC}}_{M^nCFW^n X^n})\nonumber\\[1mm]
&  \overset{\mathclap{\smash{(b)}}}{\leq}  \mathbb V(P^{\text{RB}}_{M^n C F} P^{\text{RB}}_{W^n|F} P^{\text{RB}}_{X^n| W^n M^n C}, \nonumber \\ & \phantom{\leq \mathbb V(P } Q_{M^n} Q_{C} Q_{F} P^{\text{RB}}_{W^n|F} P^{\text{RB}}_{X^n| W^n M^n C})\nonumber\\
&\,\,\,\,\,+ \mathbb V(Q_{M^n} Q_{C} Q_{F} P^{\text{RB}}_{W^n|F} P^{\text{RB}}_{X^n| W^n M^n C}, P^{\text{RC}}_{M^n C F W^nX^n }) \nonumber\\[1mm]
 &\overset{\mathclap{\smash{(c)}}}{=}  \mathbb V(P^{\text{RB}}_{M^n C F} P^{\text{RB}}_{W^n|F} P^{\text{RB}}_{X^n| W^n M^n C},\nonumber \\ 
&\phantom{\leq  \mathbb V(P } Q_{M^n} Q_{C} Q_{F} P^{\text{RB}}_{W^n|F} P^{\text{RB}}_{X^n| W^n M^n C})\nonumber\\[1mm]
 &\overset{\mathclap{\smash{(d)}}}{=}  \mathbb V(P^{\text{RB}}_{M^n C F}, Q_{M^n} Q_{C} Q_{F} )\label{tv1}
\end{align}}where $(a)$ comes from Lemma~\ref{tv prop}(\ref{cuff17}) since $\bar P_{Y^n Z^n |X^n} $ appears in both~\eqref{rb1} and~\eqref{rc1}, $(b)$ is the triangle inequality and $(c)$ comes from the definition of~\eqref{rc1}, and $(d)$ from Lemma~\ref{tv prop}(\ref{cuff17}). 

Then, to prove that~\eqref{tv1} vanishes, we need to impose rate conditions such that message $M^n $, codebook $C$ and the extra randomness $F$ are generated almost uniform and almost independent of each others. Thus, we apply Theorem~\ref{th1y} with $N=2$, $A_{\llbracket 1, 2\rrbracket}= ((X,X), W)$, $\varphi =( (\varphi_{M}, \varphi_C), \varphi_F )$, $C_{\llbracket 1, 2\rrbracket}= ((M^n,C),F)$, and $B= \emptyset$. If
{\allowdisplaybreaks
\begin{subequations}
\begin{align}
&R_F<H(W),\\
&R_{C}+ R_{M} <H(X),\label{unif 1a}\\
&R_{C}+ R_{M}+R_F <H(WX),
\label{unif 1b}
\end{align}\end{subequations}}there exists a fixed binning $\varphi^*=((\varphi_{M}^*, \varphi_C^*), \varphi_F^* ) $ such that 
$\mathbb V (P^{\text{RB},\varphi^* }_{M^n C F}, Q_{M^n} Q_{C} Q_{F})$ vanishes. 
Hence, we have \begin{equation}\label{tv3}
\mathbb V (P^{\text{RB},\varphi^*}_{M^n C F W^nX^nY^n Z^n}, P^{\text{RC}}_{M^nCFW^n X^nY^n Z^n})\to 0.
\end{equation}
From now on, we consider the binning as fixed, and we write $P^{\text{RB} }$ to simplify the notation.

\vspace{2mm}
\subsubsection{Reliable communication problem}\label{subsec: reliable communication}

Now, we want the decoder to be able to reconstruct the message. 
Thus, we apply Lemma~\ref{lem2y} to $N=2$, $\mathcal S=\{2\}$ and $\mathcal S=\emptyset$, $A_{1}= X$, $A_{2}= W$, $\varphi =( \varphi_C, \varphi_F )$, $C_{1}= C$, $C_{2}= F$, and $B=Y$. If 
 {\allowdisplaybreaks
 \begin{subequations}
 \begin{align}
 & R_{C}+ R_{F} > H(XW|Y),\label{reliability 1a}\\
 & R_{C} > H(X|WY), \label{reliability 1b}
 \end{align}\end{subequations}}the decoder recovers an estimate of the input sequence $\hat X^n$ from the binnings $(C,F)$ and side information at the decoder $Y^n$ via the Slepian--Wolf decoder with distribution $P^{\text{SW}}_{\hat X^n| CFY^n}$. Then $\hat{M}^n=\varphi_{M^n}(\hat X^n)$ is reconstructed reliably. The choice of the decoder defines the following complete joint random binning and random coding distributions:
 {\allowdisplaybreaks
 \begin{subequations}\begin{align}
& P^{\text{RB}}_{M^n C F W^nX^nY^n Z^n} P^{\text{SW}}_{\hat X^n| CFY^n} \mathds 1 \{\varphi_{M}(\hat X^n)\!=\! M^n \},\label{rb final}\\
&  P^{\text{RC}}_{M^n C F W^nX^nY^n Z^n} P^{\text{SW}}_{\hat X^n| CFY^n} \mathds 1 \{\varphi_{M}(\hat X^n)\!=\!  M^n \},\label{rc final}
\end{align}\end{subequations} }where the first part of both schemes are defined in~\eqref{rb1} and~\eqref{rc1} respectively. Moreover, since the decoder is the same in both distributions, Lemma~\ref{tv prop}(\ref{cuff17}) and~\eqref{tv3} imply
\begin{equation}\label{tv4}
\mathbb V (P^{\text{RB}}_{M^n C F W^nX^nY^n Z^n \hat X^n}, P^{\text{RC}}_{M^n C F W^nX^nY^n Z^n \hat X^n} )\! \to \! 0.
\end{equation}

\vspace{2mm}
\subsubsection{Reducing the rate of common randomness -- Remote strong coordination of $Z^n$}\label{subsec: reduce} 
With rate conditions~\eqref{unif 1a},~\eqref{unif 1b},~\eqref{reliability 1a}, and~\eqref{reliability 1b} we ensure reliability and strong coordination of the whole sequence $(M^n, C, F, W^n, X^n, Y^n, Z^n)$. 
First, observe that 
Lemma~\ref{tv prop}(\ref{cuff16}) and~\eqref{tv4} imply 
\begin{equation}\label{tv5}
\mathbb V (P^{\text{RB}}_{M^n F Z^n} , P^{\text{RC}}_{M^n F Z^n}) \to 0.
\end{equation}

As in~\cite{yassaee2014achievability}, we would like to reduce the amount of common randomness by having the two nodes agree on an instance $F=\mathbf f$. 
The first step, is to ensure that there exists a binning such that  $F$ is generated almost uniformly and independent of $(M^n, Z^n)$. 
To do this, we apply Corollary~\ref{cor1y} with $N=2$, $ A_1=W$, $A_2=X$, $B=Z$, and $\mathcal V=\emptyset$. If
{\allowdisplaybreaks
\begin{subequations}
\begin{align}
& R_{F}+R_{M} <H(WX|Z),\label{rate mza}\\
&R_{F} <H(W|Z),\label{rate mzb}
\end{align}\end{subequations}}we obtain
\begin{equation}\label{tv6}
\mathbb V (P^{\text{RB}}_{M^n F Z^n} , Q_F P^{\text{RB}}_{M^n Z^n}) \to 0.
\end{equation}Combining the triangle inequality with~\eqref{tv5} and~\eqref{tv6}, we have
{\allowdisplaybreaks\begin{align}
& \mathbb V (Q_F P^{\text{RC}}_{M^n Z^n}, Q_F P^{\text{RB}}_{M^n Z^n}) \leq \mathbb V (Q_F P^{\text{RC}}_{M^n Z^n} , P^{\text{RB}}_{M^n F Z^n}) \nonumber\\
&\qquad+\mathbb V (P^{\text{RB}}_{M^n F Z^n} , Q_F P^{\text{RB}}_{M^n Z^n}) \to 0.\label{tv7}
\end{align}}The second and final step is to use Lemma~\ref{tv prop}(\ref{lem4y}), which ensures that, given~\eqref{tv7}, there exists an instance $\mathbf f \in \llbracket 1, 2^{n R_{F}}\rrbracket$ such that  
\begin{equation}\label{tv8}
\mathbb V (P^{\text{RC}}_{M^n Z^n|F=\mathbf f },  P^{\text{RB}}_{M^n Z^n|F=\mathbf f})\to 0.
\end{equation}

\vspace{2mm}
\begin{rem}[Controlling the joint distribution--remotely coordinating $Z^n$]
Observe that controlling the joint distribution $\bar P_{Z^n M^n}$ is equivalent to controlling the conditional distribution $\bar P_{Z^n|M^n}$,  since  $M^n$ is generated uniformly by assumption. Thus, we are \emph{remotely} coordinating $Z^n$.
\end{rem}

\vspace{2mm}
\begin{rem}[One binning for both conditions]
We take $(\varphi^{*}_{F}, \varphi^{*}_{M})$ $=$ $( \varphi_{F}',  \varphi_{M}')$, which works for the conditions of both Section~\ref{subsec: joint strong coord} and Section~\ref{subsec: reduce}. For a  thorough discussion on the existence of such binning, see \cite[Remark 3.7]{cervia2018thesis}.
\end{rem}

\vspace{2mm}

\begin{rem}[Stochasticity of encoder and decoder]
Observe that, if   instead of the stochastic function $g_n$, we consider the deterministic decoder $\tilde g_n$, that exploits external randomness $U$, we can always represent discrete stochastic decoders as discrete deterministic decoders with auxiliary randomness $U$.  
Each realization $\mathbf u $ of $U$ gives a deterministic decoder, and the average over all $\mathbf u $ is equal to $p_e=\mathbb P \{ M^n \! \!   \neq  \! \hat{M}^n\}$, since
\begin{equation*}
p_e  \! =\! \mathbb P \{ M^n \!   \neq  \! \hat{M}^n\}\!=\! \mathbb E_{U} \!  \left[  \mathbb P \{ M^n  \!  \neq \!  \hat{M}^n |U\}   \right]\! = \! \mathbb E_{U} \!  \left[ \, p_e(U)\right].
\end{equation*}Hence, there exists at least one choice  $\mathbf u^{\star}$ for which $p_e(\mathbf u^{\star}) \leq p_e$.   Since the choice of a deterministic decoder only concerns reliable communication and not approximating the target distribution, the decoder can be deterministic without loss of generality. Note that we cannot apply the same reasoning to the encoder:   by assuming that the encoder is deterministic, we would restrict the choice of distributions $\bar P_{Z|M}$ that can be coordinated. Therefore, we will not achieve the whole coordination region.
\end{rem}


\vspace{1.5mm}
\subsubsection{Rate conditions}\label{subsec: reduce} We retrieve the rate conditions of~\eqref{region gen} by performing Fourier--Motzkin elimination with respect to $R_F$.
More precisely, for the message rate we have
{\allowdisplaybreaks
\begin{subequations}\begin{align}
 R_{M}& \overset{\mathclap{\smash{(a)}}}{<}  H(WX|Z), \label{condition M1}\\
 R_{M}& \overset{\mathclap{\smash{(b)}}}{<}  H(X)+H(W|X)- (R_{C}+ R_{F} )\nonumber\\
 &\overset{\mathclap{\smash{(c)}}}{<} H(X)+H(W|X)- H(XW|Y)\nonumber\\
 &=H(X)- H(X|Y)+H(W|X)- H(W|XY)\nonumber\\
 &=I(X;Y)+I(W;Y|X)\overset{\mathclap{\smash{(d)}}}{=} I(X;Y),\label{condition M2}\\
 R_{M}& \overset{\mathclap{\smash{(e)}}}{<}  H(X)- R_{C}\nonumber\\
 & \overset{\mathclap{\smash{(f)}}}{<} H(X)-H(X|WY)=I(X;WY),\label{condition M3}
\end{align}\end{subequations} }where $(a)$ follows from~\eqref{rate mza}, $(b)$ is condition~\eqref{unif 1b}, $(c)$ comes from~\eqref{reliability 1a} and $(d)$ from the Markov Chain $W-X-Y$. Finally, $(e)$ is condition~\eqref{unif 1a} and $(f)$ follows from~\eqref{reliability 1b}. Then, by combining~\eqref{condition M1},~\eqref{condition M2}, and~\eqref{condition M3}, we have
{\allowdisplaybreaks
\begin{align}
R_{M} <& \min\{H(WX|Z), I(X;Y), I(X;WY) \}\nonumber\\
=& \min\{H(WX|Z), I(X;Y) \}.\label{condition M4}
\end{align}} 
For the rate of common randomness, we have
{\allowdisplaybreaks
\begin{subequations}\begin{align}
 R_{C} &\overset{\mathclap{\smash{(g)}}}{>}   H(XW|Y)-R_F \nonumber\\
 &\overset{\mathclap{\smash{(h)}}}{>}  H(XW|Y)- H(W|Z)\nonumber\\
 &= H(X|WY)+ H(W|Y)-H(W|Z)\nonumber\\
 &= H(X|WY)+  I(W;Z)- I(W;Y)\label{condition C1}\\
  R_{C} &\overset{\mathclap{\smash{(i)}}}{>}   H(X|WY), \label{condition C2}
\end{align}\end{subequations} }where $(g)$ is condition~\eqref{reliability 1a}, $(h)$ follows from~\eqref{rate mzb}, and $(i)$ is condition~\eqref{reliability 1b}.  Observe that
~\eqref{condition C1} and~\eqref{condition C2} imply
{\allowdisplaybreaks
\begin{align}
R_{C} >\max\{0, I(W;Z)- I(W;Y)\}.\label{condition C3}
\end{align}}


\subsection{Achievability Proof of Proposition~\ref{secrecy region}}\label{sec: ac proof secrecy region}
The achievability follows from the proof of Theorem~\ref{theogeneral} by adding to rate constraints~\eqref{unif 1a},~\eqref{unif 1b},~\eqref{reliability 1a},~\eqref{reliability 1b},~\eqref{rate mza} and~\eqref{rate mzb} the conditions to ensure that the message is independent of $Z^n$. Thus, we apply Theorem~\ref{th1y} with $N=1$, $A_1=X$, $B=Z$. If
\begin{equation}\label{rate mz secrecy}
R_{M} <H(X|Z),
\end{equation}the distance $\mathbb V (Q_{M^n} P^{\text{RB}}_{Z^n} ,  P^{\text{RB}}_{M^n Z^n}) $ vanishes, and therefore
\begin{equation}\label{tv9}
\mathbb V (Q_{M^n} P^{\text{RB}}_{Z^n} ,  P^{\text{RC}}_{M^n Z^n}) \to 0.
\end{equation}

\vspace{2mm}
\subsection{Achievability Proof of Proposition~\ref{control z}} \label{sec: ac proof control z}
The first three steps are the same as in Sections~\ref{subsec: def random schemes},~\ref{subsec: joint strong coord},  and~\ref{subsec: reliable communication}. In particular, the rate constraints~\eqref{unif 1a},~\eqref{unif 1b},~\eqref{reliability 1a}, and~\eqref{reliability 1b} hold. Moreover, by Lemma~\ref{tv prop}(\ref{cuff16}), we have
\begin{equation}\label{tv10}
\mathbb V(P^{\text{RC}}_{F Z^n}, P^{\text{RB}}_{F Z^n})\to 0.
\end{equation}However, when reducing the rate of common randomness, we only want to control the marginal distribution $\bar P_{Z^n}$. Hence, we apply Theorem~\ref{th1y} with $N=1$, $A_1=W$, $B=Z$. If
\begin{equation}\label{rate mz marginal}
R_{F} <H(W|Z),
\end{equation}the distance $\mathbb V (Q_{F} P^{\text{RB}}_{Z^n} ,  P^{\text{RB}}_{FZ^n}) $ vanishes, and therefore
\begin{equation}\label{tv11}
\mathbb V (Q_{F} P^{\text{RB}}_{Z^n} ,  P^{\text{RC}}_{F Z^n}) \to 0.
\end{equation} 

\vspace{2mm}
\section{Converse}\label{section converse}
The converse proofs use classical tools such as properties of mutual information and entropy, Fano's Inequality, and the Csisz\'ar Sum Identity. Moreover, we will use the following technical result, well-suited for converses in coordination problems.
\begin{lem}[$\mbox{\cite[Lemma 5]{Cervia2017}}$]\label{lemmit}
Let $\bar P_{A}^{\otimes n}$ be i.i.d., and $P_{A^{n}}$  be such that $\mathbb V(P_{A^{n}}, \bar P_{A}^{\otimes n})$ vanishes.
Then, we have that 
\begin{equation*}
\sum\nolimits_{t=1}^{n} I(A_t;A_{\sim t}) \leq n\, \varepsilon. 
\end{equation*}
\end{lem}

\vspace{2mm}
\subsection{Converse Proof of Theorem~\ref{theogeneral}}
Consider a code $(f_n,g_n)$ that induces a distribution $P_{Z^{n}|M^n}$ on the actions that is $\varepsilon$-close in total variation distance to the i.i.d. distribution $\bar{P}_{Z|M}^{\otimes n}$ and such that $\mathbb P \{ M^n \neq \hat{M}^n\} \leq \varepsilon$.
Furthermore, let the random variable $T$ be uniformly distributed over the set $\llbracket 1,n\rrbracket$ and independent of the variables $(X^{n}, Y^{n}, Z^n,  C, M^n)$.

\vspace{2mm}
\subsubsection{Rate of  common randomness}\label{rate of cr}
We have
{\allowdisplaybreaks\begin{align}
& n R_{C} = H(C)\geq I(C; M^n Z^n)\nonumber\\
&  = \sum_{t=1}^n I(C; M_t Z_t|Z^{t-1} M^{t-1}) \geq \sum_{t=1}^n I(C; Z_t|Z^{t-1} M^{t-1})\nonumber\\
&  \geq \sum_{t=1}^n \left[ I(C;  Z_t|Z^{t-1} M^{t-1}) - I(C; Y_t|Y_{t+1}^n M^{t-1})\right]\nonumber\\
&=   \sum_{t=1}^n   \left[ I(C Y_{t+1}^{n}; Z_t|Z^{t-1}M^{t-1})  \right. \nonumber\\
&\qquad\qquad   \left.-  I(Y_{t+1}^{n}; Z_t|Z^{t-1}M^{t-1} C) \right]\nonumber\\
&\quad-  \sum_{t=1}^n   \left[ I(C Z^{t-1}; Y_t|Y_{t+1}^n M^{t-1})  \right.\nonumber\\
& \qquad\qquad  -  \left. I(Z^{t-1}; Y _t|Y_{t+1}^n M^{t-1} C)\right] \nonumber\\
& \overset{\mathclap{\smash{(a)}}}{=}   \sum_{t=1}^n   \left[ I(C Y_{t+1}^{n}; Z_t|Z^{t-1}M^{t-1})  \right. \nonumber\\
&  \qquad\qquad\left. -   I(C Z^{t-1}; Y_t|Y_{t+1}^n M^{t-1})\right]\nonumber\\
&  \overset{\mathclap{\smash{(b)}}}{\geq}   \sum_{t=1}^n   \left[ I(C Z^{t-1 }Y_{t+1}^{n} M^{t-1 };  Z_t) \right. \nonumber\\
& \qquad\qquad - \left. I(C Z^{t-1};  Y_t|Y_{t+1}^n M^{t-1})\right] - n \varepsilon\nonumber\\
&  \geq   \sum_{t=1}^n  \left[ I(C Z^{t-1}Y_{t+1}^{n} M^{t-1}; Z_t) \right.\nonumber\\
&\qquad\qquad \left. - I(C Z^{t-1 } Y_{t+1}^n M^{t-1 }; Y_t) \right] - n \varepsilon\nonumber\\
&  \overset{\mathclap{\smash{(c)}}}{=}   \sum_{t=1}^n\left[ I(W_t ; Z_t) -  I(W_t ; Y_t) \right]-n \varepsilon\nonumber\\
&  = n I(W_T ; Z_T| T) - n I(W_T ; Y_T| T)-n \varepsilon\nonumber\\
&  \overset{\mathclap{\smash{(d)}}}{\geq}  n I(W_T T; Z_T) - n I(W_T T; Y_T)-2n \varepsilon\nonumber\\
&  \overset{\mathclap{\smash{(e)}}}{=}   n I(W ; Z) - n I(W ; Y)-2n \varepsilon\label{rate C gc}
\end{align}}where $(a)$ comes from the Csisz\'ar Sum Identity. To prove  $(b)$, observe that 
{\allowdisplaybreaks
\begin{align}
&  I(C Y_{t+1}^{n}Z^{t-1}M^{t-1}; Z_t)-I(C Y_{t+1}^{n}; Z_t|Z^{t-1}M^{t-1})\nonumber\\
&\quad =I(Z_t;Z^{t-1}M^{t-1})\leq I(M_t Z_t; Z_{\sim t} M_{\sim t}) \leq \varepsilon
\end{align}}from Lemma~\ref{lemmit} and assumption \eqref{coordz}.
Then, $(c)$  follows from the identification $W_t=(C, Z^{t-1}, Y_{t+1}^n, M^{t-1})$, 
$(d)$ from the fact that $I(Z_T;T)\leq \varepsilon$ since $Z^n$ is almost i.i.d. by assumption \eqref{coordz} and $ I(W_T ; Y_T| T)\leq I(W_T T; Y_T) $. Finally,  $(e)$ comes from the identification $  W=(W_T,T)$.

\vspace{1mm}

\subsubsection{Rate of the message}

\paragraph{$R_{M} < \max_{P_X} I(X;Y) $}\label{rateR}It follows from Fano's Inequality, since the probability of error tends to zero.
For more details, see~\cite{elgamal2011nit}.

\paragraph{$R_{M} < \max_{P_X} H(XW|Z)$}\label{rateH(WX|Z)} We have 
 {\allowdisplaybreaks\begin{align}
nR_{M}& = H(M^n) = \sum_{t=1}^{n} H(M_t|M^{t-1}) \overset{\mathclap{\smash{(a)}}}{=}   \sum_{t=1}^{n} H(M_t)   \nonumber\\
& \overset{\mathclap{\smash{(b)}}}{\leq}  \sum_{t=1}^{n} H(M^{t-1})\nonumber \overset{\mathclap{\smash{(c)}}}{\leq}  \sum_{t=1}^{n} H(M^{t-1}Z^{t-1}|Z_t)+n \varepsilon\nonumber \\
& \leq  \sum_{t=1}^{n} H(M^{t-1}Z^{t-1} Y_{t+1}^n C X_t|Z_t)+n \varepsilon\nonumber\\
& \overset{\mathclap{\smash{(d)}}}{=} \sum_{t=1}^{n} H(W_t X_t|Z_t)+n \varepsilon = n H(W_T X_T|Z_T T)\! +\! n \varepsilon\nonumber\\
& \leq  n H(W_T X_T T|Z_T)\! +\! n \varepsilon \overset{\mathclap{\smash{(e)}}}{=}  n H(WX|Z)\! +\! n \varepsilon\label{converseR2a} 
\end{align}}where $(a)$ comes  from the fact that $M^n$ is  i.i.d., and $(b)$ holds  from the same reason if $n>2$. Then, $(c)$ follows from
 {\allowdisplaybreaks\begin{align}
&H(M^{t-1}) -H(M^{t-1} Z^{t-1} |Z_t)\nonumber\\
& \leq  H(M^{t-1} Z^{t-1}) - H(M^{t-1} Z^{t-1} |Z_t)\nonumber\\
&=  I(Z_t; Z^{t-1} M^{t-1}) \leq   I(M_t Z_t; Z_{\sim t}M_{\sim t}) \leq \varepsilon
\end{align}}by Lemma~\ref{lemmit} and assumption~\eqref{coordz}.
Finally,  in steps $(d)$ and $(e)$ we identify $W_t=(C, Z^{t-1}, Y_{t+1}^n, M^{t-1})$, $W=(W_T,T)$. Observe that $W_t$ defined in this way verifies the Markov Chain $W_t-X_t-(Y_t,Z_t)$.

 \vspace{1mm}
\paragraph{Cardinality bound}
The cardinality bound on $\mathcal{W}$  is a consequence of the Fenchel--Eggleston--Carath\'eodory Theorem~\cite[Appendix C]{elgamal2011nit}. The proof is omitted.

\subsection{Converse Proof of Proposition~\ref{secrecy region}}\label{converse proof secrecy region}
\subsubsection{Rate of  common randomness}\label{rate cr 2}
See Section~\ref{rate of cr}.
 \vspace{1mm}

\subsubsection{Rate of the message}\label{rateR2} 
 \vspace{3mm}
 \paragraph{$R_{M} < \max_{P_X} I(X;Y) $} See Section \ref{rateR}
 \vspace{1mm}
\paragraph{$R_{M} < \max_{P_X} H(X|Z)$} We have 
 {\allowdisplaybreaks\begin{align}
nR_{M} &=  H(M^n)  \nonumber\\
 & \overset{\mathclap{\smash{(a)}}}{=} H(M^n|Z^n) + n\varepsilon \nonumber\\
 & \leq H(M^{n\!} X^n|Z^n) \!+\! n\varepsilon  \nonumber \\
 & = H(X^n|Z^n) \!+\! H(M^n |X^{n\!} Z^n)\!+\! n\varepsilon  \nonumber\\
  &\leq  H(X^n|Z^n) + H(M^n |X^n) + n\varepsilon \nonumber\\
 & \overset{\mathclap{\smash{(b)}}}{\leq}  H(X^n|Z^n) +3n  \varepsilon \nonumber\\
 & =  \! \sum_{t=1}^n \!H(X_t|X^{t-1 \!}Z^n)+ 3n \varepsilon \nonumber\\
&\leq \sum_{t=1}^n \!H(X_t|Z_t) +3n \varepsilon\nonumber\\
& =  n H(X_T|Z_T, T) + 3n\varepsilon \nonumber\\
 &  \leq  n H(X|Z)+ 3n \varepsilon\label{converseR2} 
\end{align}}where $(a)$ comes from~\eqref{secrecyz}.
To prove  $(b)$, observe that 
 {\allowdisplaybreaks\begin{align}
 \MoveEqLeft[2]H(M^n|X^n)  \leq H(M^n \hat{M}^n|X^n)   \nonumber\\[1mm]
 &=  H(\hat{M}^n |X^n) + H( M^n |X^n \hat{M}^n)  \nonumber\\[1mm]
  &   \overset{\mathclap{\smash{(c)}}}{\leq}   H(\hat{M}^n |X^n) +n \varepsilon  \overset{\mathclap{\smash{(d)}}}{\leq}   H(\hat{M}^n | M^n C)+n \varepsilon \nonumber\\[1mm]
&\leq  H(\hat{M}^n | M^n )+n\varepsilon  \overset{\mathclap{\smash{(e)}}}{\leq}  2 n \varepsilon \label{converseR2a part 2} 
\end{align}}where $(c)$ comes from the fact that $H( M^n |X^n \hat{M}^n)  \leq  H({M^n} | \hat{M}^n) \leq n \varepsilon$ by Fano's Inequality  since the probability of $M^n$ being different from $\hat{M}^n$ tends to zero. Then, because of the Markov chain ${M^n}C-X^n-Y^n- \hat{M}^n$, $(d)$ follows from the data processing inequality. Finally, $(e)$ comes from  Fano's Inequality.

\subsection{Converse Proof of Proposition~\ref{control z}}\label{converse proof control z}

 \vspace{2mm}
\subsubsection{Rate of  common randomness}\label{rate cr 3}
The proof is similar to the one in Section~\ref{rate of cr}, but since the assumptions are slightly different, we need a different identification for the auxiliary random variable $W$. More precisely, we have
{\allowdisplaybreaks\begin{align}
&nR_{C} = H(C)\geq I(C;Z^n)\geq I(C;Z^n)- I(C;Y^n)\nonumber\\
& = \sum_{t=1}^n I(C; Z_t|Z^{t-1}) - \sum_{t=1}^n I(C; Y_t|Y_{t+1}^n)\nonumber\\
& \overset{\mathclap{\smash{(a)}}}{=}  \sum_{t=1}^n I(C Y_{t+1}^{n}; Z_t|Z^{t-1}) - \sum_{t=1}^n I(C Z^{t-1}; Y_t|Y_{t+1}^n)\nonumber\\
& \overset{\mathclap{\smash{(b)}}}{\geq}  \sum_{t=1}^n I(C Z^{t-1}Y_{t+1}^{n} ; Z_t) - \sum_{t=1}^n I(C Z^{t-1}; Y_t|Y_{t+1}^n)-n \varepsilon\nonumber\\
& \geq \sum_{t=1}^n I(C Z^{t-1}Y_{t+1}^{n} ; Z_t) - \sum_{t=1}^n I(C Z^{t-1} Y_{t+1}^n; Y_t)-n \varepsilon\nonumber\\
& \overset{\mathclap{\smash{(c)}}}{=}  \sum_{t=1}^n I(W_t ; Z_t) - \sum_{t=1}^n I(W_t ; Y_t)-n \varepsilon\nonumber\\
& = n I(W_T ; Z_T| T) - n I(W_T ; Y_T| T)-n \varepsilon\nonumber\\
& \overset{\mathclap{\smash{(d)}}}{\geq}  n I(W_T T; Z_T) - n I(W_T T; Y_T)-2n \varepsilon\nonumber\\
& \overset{\mathclap{\smash{(e)}}}{=}   n I(W ; Z) - n I(W ; Y)-2n \varepsilon\label{rate C gc}
\end{align}}where $(a)$ comes from the Csisz\'ar Sum Identity. To prove  $(b)$, observe that 
{\allowdisplaybreaks
\begin{align}
&  I(C Y_{t+1}^{n}Z^{t-1}; Z_t)-I(C Y_{t+1}^{n}; Z_t|Z^{t-1})\nonumber\\
&\qquad =I(Z_t;Z^{t-1})\leq I(M_t Z_t; Z_{\sim t} M_{\sim t}) \leq \varepsilon
\end{align}}from Lemma~\ref{lemmit} and assumption \eqref{coordz2}.
Then, $(c)$  follows from the identification $W_t=(C, Z^{t-1}, Y_{t+1}^n)$, 
$(d)$ from the fact that $I(Z_T;T)\leq \varepsilon$ since $Z^n$ is almost i.i.d. by assumption \eqref{coordz} and $ I(W_T ; Y_T| T)\leq I(W_T T; Y_T) $. Finally,  $(e)$ comes from the identification $  W=(W_T,T)$.
  
 \vspace{2mm}  
\subsubsection{Rate of the message}\label{rateR3} 
See Section~\ref{rateR}.


\begin{small}
\bibliographystyle{IEEEtran}
\bibliography{mybib}
\end{small}

\end{document}